\documentclass{article}

\PassOptionsToPackage{numbers, sort&compress}{natbib}
\usepackage[dblblindworkshop, final]{neurips_2026}
\workshoptitle{Representations for the Physical Sciences}

\usepackage[utf8]{inputenc} % allow utf-8 input
\usepackage[T1]{fontenc}    % use 8-bit T1 fonts
\usepackage[pdftex]{hyperref}       % hyperlinks
\usepackage{url}            % simple URL typesetting
\usepackage{booktabs}       % professional-quality tables
\usepackage{amsfonts}       % blackboard math symbols
\usepackage{nicefrac}       % compact symbols for 1/2, etc.
\usepackage{microtype}      % microtypography
\usepackage[dvipsnames, table]{xcolor}         % colors
\usepackage{graphicx}
\usepackage{stackengine}
\usepackage{xspace}
\usepackage[justification=raggedright]{subcaption}
\usepackage{placeins}
\usepackage{amsmath}
\usepackage{bbm}
\usepackage{siunitx}
\usepackage[normalem]{ulem}

\definecolor{blue1}{HTML}{1065AB} % blue1 % R 016 G 101 B 171
\definecolor{red2}{HTML}{B31529} % red2   % R 179 G 021 B 041

\hypersetup{
    colorlinks=true, % Enable colored links instead of boxes
    linkcolor=blue1,
    citecolor=blue1,
    urlcolor=blue1,
}

\title{ARENA: Active Region attENtion Aggregation for interpretable solar flare prediction}

\author{%
  Luis Bauer\\
  MPI for Intelligent Systems, Tübingen\\
  University of Tübingen
  \\
  \And
  Annalena~Kofler\thanks{Corresponding author: \texttt{annalena.kofler@tuebingen.mpg.de}}\\
  MPI for Intelligent Systems, Tübingen\\
  MPI for Gravitational Physics, Potsdam\\
  \AND
  Andrés Asensio Ramos \\
  Instituto de Astrofísica de Canarias \\
  Universidad de La Laguna \\
  \And
  Aparna Venkataramanasastry \\
  New Mexico State University
  \AND
  Valentín Martínez Pillet \\
  Instituto de Astrofísica de Canarias \\
  Universidad de La Laguna 
  \And
  Bernhard~Schölkopf \\
  MPI for Intelligent Systems, Tübingen\\
  ELLIS Institute T\"ubingen \\
}

\begin{document}

\maketitle

\begin{abstract}
    Predicting solar flares is a challenging space-weather forecasting problem with potentially significant impacts on Earth's upper atmosphere, including disruptions to radio communications.
    We propose \textbf{Active Region attENtion Aggregation~(ARENA)}, a neural network architecture that predicts solar flares from a variable number of active-region cutouts while learning to identify the region most relevant to the prediction.
   ARENA is trained using multiple-instance learning under weak supervision, requiring only a flare/no-flare label for the observation and no labels specifying the flare origin.
    The learned attention mechanism therefore provides a means of localizing the active region driving the prediction without explicit origin annotations.
    ARENA achieves a heidke skill score of 0.45 which is comparable to other flare prediction approaches.
    These results demonstrate that active-region-based aggregation can provide an efficient and interpretable approach to solar flare prediction.
    The code is available at \url{https://github.com/LuisBauer22/ARENA}.
\end{abstract}

\section{Introduction}

\paragraph{Solar flare forecasting.} Solar flares are transient releases of magnetic energy in the solar atmosphere that produce intense electromagnetic radiation. 
These events are associated with active regions~(ARs), localized concentrations of strong and complex magnetic fields on the solar surface~\citep{Overview_of_solar_flares}. 
Radiation from the most energetic flares can perturb Earth's ionosphere and affect radio communication and space-based technologies, making reliable flare forecasting an important problem in space-weather research.
A central challenge in solar flare forecasting is the spatial scale at which the problem is formulated: 
A flare is observed as a global event, but its magnetic origin is localized to an AR. 
Consequently, two related questions arise: \emph{Will a flare occur within \SI{24}{\hour}, and which AR is responsible for it?} 
The first question can be addressed using event-level flare catalogs, whereas the second requires associating individual flares with their source regions. 
Most existing data-driven forecasting approaches focus on the former, using either full-disk observations~\citep{Pandey_flares_2024_full_disk, roy2025surya, Yan_2024_full_disk_flares} or individual ARs~\citep{Li_flares_2023, Sun_flares_2022, Tang_flares_2021, Huang_flares_2018} as input.
In contrast, combining~ARs present in an observation provides a natural representation of the solar disk as a variable-sized set of physically meaningful regions.

\paragraph{Multiple-instance learning for active region identification.} Representations combining ARs introduce a mismatch between the available data and the desired labels. 
For each observation, it is straightforward to obtain a global label indicating whether a flare occurs within the subsequent~\SI{24}{\hour}. 
Determining which of the~ARs present at the observation time produces the subsequent flare, however, requires additional event-to-region association based on expert knowledge. 
Such origin annotations are consequently substantially more difficult to obtain and are available only for limited datasets~\citep{SWAN_SF_flare_AR_dataset, Boucheron_2023_solar_AR_dataset}.
We therefore seek to learn from the readily available global flare labels while retaining the ability to identify the~AR driving the prediction.
This setting corresponds naturally to \emph{multiple-instance learning}~(MIL~\citep{Carbonneau_2018_MIL_Overview}). 
In MIL, a bag contains multiple instances but is associated only with a bag-level label. 
This paradigm has been particularly successful in computational pathology, where a high-resolution whole-slide image may be labeled according to the presence of a disease, such as cancer, while identifying the diagnostically relevant tissue regions requires costly expert annotation~\citep{ilse2018_abmil, shao2021_transmil, cui2023bayesmil, zhu2024dgrmil}. 
We formulate solar flare forecasting analogously: 
An observation constitutes a bag, its associated AR~cutouts are the instances, and the flare/no-flare label is assigned to the bag. 
The number of ARs can vary between observations, requiring an aggregation mechanism that is invariant to the number and ordering of instances. 
Importantly, information about the AR associated with a flare is not available during training.

\begin{figure*}[t]
    \centering
    \includegraphics[width=\textwidth]{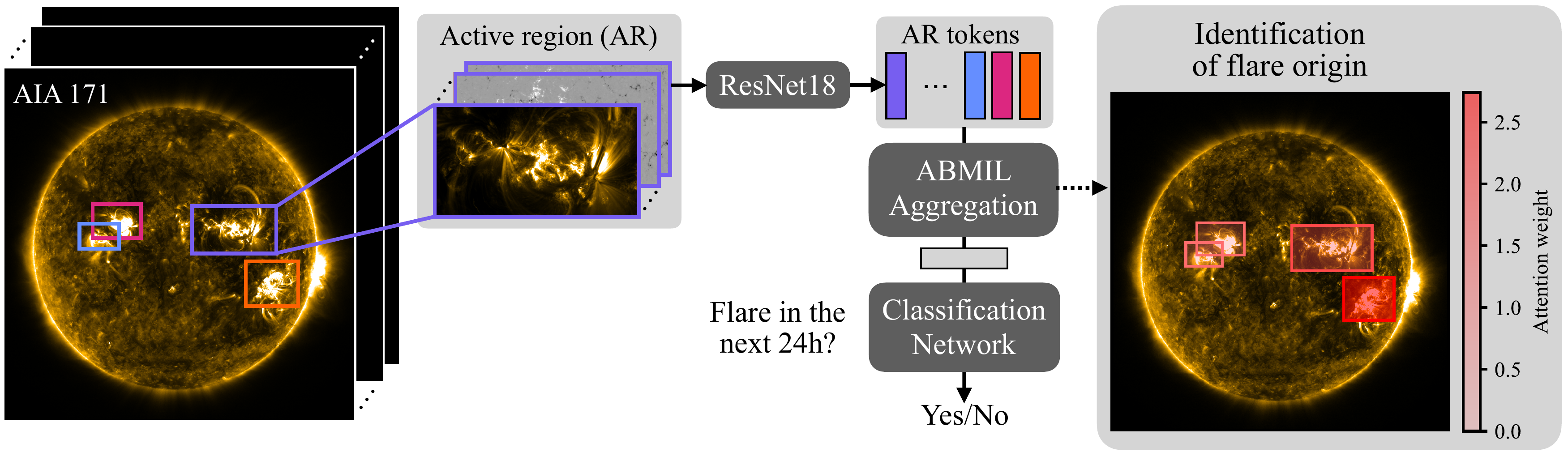}
    \caption{\label{fig: architecture} Overview of the ARENA architecture. Each AR across all channels is extracted, rescaled, and provided to a ResNet-18 which compresses the image into a token vector.
    For one time point, all AR tokens are concatenated and passed through an ABMIL aggregator, resulting in a single summary vector.
    This vector is processed by a classification head which predicts whether a flare is happening in the next~\SI{24}{\hour}.
    At inference, the attention weight can be employed to identify the flare-originating~AR.
    }
    \label{fig:architecture}
\end{figure*}

\paragraph{Contributions.}
% aka What do we REALLY do ourselves?
(1)~We propose \textbf{ARENA (Active Region attENtion Aggregation)}, an attention-based neural network architecture that represents solar observations as variable-sized sets of physically motivated active region instances, illustrated in Figure \ref{fig: architecture}.
(2)~We formulate solar flare forecasting as a MIL problem, using only observation-level flare labels while treating the associated active regions as instances. This enables training without explicit labels for the flare-producing region.
(3)~We show that the learned attention weights can identify the AR associated with the observed flare, despite the absence of origin annotations during training. We show that the ARENA model provides an interpretable localization signal while maintaining comparable flare forecasting performance.

\section{ARENA: Active Region attENtion Aggregation}
\label{sec: ARENA}
%\looseness=-1
\paragraph{MIL problem formulation.}
We formulate solar flare forecasting as a MIL~problem where each observation at time~$t$ is represented as a bag~$\mathcal{B}(t) = \{\mathbf{x}_1, ..., \mathbf{x}_{N_t}\}$ containing the active regions identified in the corresponding AR catalog.
Here, $\mathbf{x}_i$ denotes the image cutout of the $i$th~AR at time~$t$ and $N_t$ denotes the number of ARs present at that timestamp.
Consequently, the number of instances varies between observations.
Observations containing no~ARs are excluded from the dataset; the distribution of~$N_t$ is reported in Appendix~\ref{appendix:data_set}.
This AR-based data representation focuses the model on physically relevant regions and avoids processing the substantially larger full-disk images, significantly reducing memory and disk requirements.

Each bag is assigned a binary label
\begin{align}
    y_t = \begin{cases}
        1 & \exists t_\mathrm{flare}:~t < t_\mathrm{flare} \leq t+\SI{24}{\hour}~, \\
        0 & \text{otherwise}~,
    \end{cases}
\end{align}
if there is at least one M- and X-class flare within the subsequent 24 hours.
During training, only the bag-level label~$y_t$ is provided. 
For a positive bag, the active region responsible for the subsequent flare is not annotated; instead, we assume that the flare-producing AR is among the ARs present at time~$t$. 
Consequently, the model must learn which ARs are relevant to the bag-level prediction from the global flare label alone. 
This constitutes weak supervision at the AR~level and motivates an attention-based aggregation mechanism capable of handling a variable number of AR instances.

\paragraph{ARENA architecture.}
ARENA encodes each~AR independently and subsequently aggregates the resulting representations into a single bag-level representation. 
Given an observation~$\mathcal{B}(t)$\, as illustrated in Figure~\ref{fig: architecture}, each AR~cutout is mapped to an embedding~$\mathbf{h_i} = f_{\theta}\big(\mathbf{x_i}\big)$, where $f_{\theta}$ is a ResNet-18~\citep{ResNet} image encoder.
For the main ARENA model, we use attention-based multiple-instance learning (ABMIL~\citep{ilse2018_abmil}) to aggregate the AR~embeddings. 
Specifically, the attention weights are computed as
\begin{align}
    \label{eq: abmil coeff}
    \alpha_i = \frac{\exp \left[a_{\phi}\big(\mathbf{h_i}\big)\right]} {\sum_{j=1}^{N_t}\exp \left[a_{\phi}\big(\mathbf{h_j}\big)\right]}~,
\end{align}
where $a_{\phi}$ is a learned attention function.
The resulting bag representation is given by the attention-weighted sum
$\mathbf{z} = \sum_{i=1}^{N_t}\alpha_i \mathbf{h_i}$.
The exact gated attention formulation and its architectural details are provided in Appendix~\ref{appendix: architecture_details}. 
A classification head then maps~$\mathbf{z}$ to the probability of an M+~flare within the following~\SI{24}{\hour}.
Since ARENA performs the aggregation in the learned feature space, it can learn both which AR~representations are relevant and which features within these representations contribute to the bag-level prediction.

\section{Experimental Setup}
\label{sec: experimental setup}
\paragraph{Solar data.}
We use solar observations from 2010--2024 at a cadence of two~hours, with the active regions identified from the corresponding AR~catalog and represented by image cutouts~\citep{Bobra_2014_sharps}. 
Flare labels are obtained from the NOAA Geostationary Operational Environmental Satellite X-ray sensor flare (GOES XRS) catalog~\citep{GOES_XRS_flare_catalog}. 
The data from 2010--2016 are used for training and validation, while 2017--2024 are held out as an independent test set. 
Within the development period, we perform five temporal validation folds, each using two months of each year for validation and a one-month temporal buffer between the training and validation periods to avoid temporal temporal leakage. 
For evaluating AR source identification, we use the subset of the flare dataset containing annotations of the AR associated with each flare~\citep{SWAN_SF_flare_AR_dataset} corresponding to a set of flares in~2017. 
Further details on data preprocessing and dataset composition are provided in Appendix~\ref{appendix:data_set}.

\paragraph{Training.}
Each model is optimized using AdamW~\citep{adammethodstochasticoptimization_kingma_2015} with a cross-entropy loss. 
All five models are subsequently evaluated on the same held-out test period from 2017--2024. 
Further details on training hyperparameters are provided in Appendix~\ref{appendix:training_details}.

\paragraph{Evaluation.}
We evaluate ARENA along two complementary dimensions: Flare forecasting and AR source identification.
For flare forecasting, we use the Heidke Skill Score~(HSS) as the primary metric  across the validation and temporally held-out test sets. 
The HSS measures the improvement of a binary classifier over a random forecast with optimal class frequencies, see Appendix~\ref{appendix: evaluation}.
To assess whether the weakly supervised aggregation limits predictive performance, we additionally compare ARENA with an instance-level classifier using the same AR encoder but trained with explicit flare-AR annotations~\citep{SWAN_SF_flare_AR_dataset}.
For this comparison, we report additional classification metrics to characterize differences between direct AR-level and bag-level prediction.
Details of the evaluation protocol and baseline comparisons are provided in Appendix~\ref{appendix: architecture_details} and Appendix~\ref{appendix: additional results}, respectively.
\\
For AR~source identification, we rank ARs within each positive bag according to their attention weights and quantify source-AR localization using top-$k$ accuracy~(Acc@$k$). 
This measures the fraction of observations for which the annotated flare-producing~AR is among the $k$~highest-attention~ARs:
\begin{align}
    \mathrm{Acc}@k = \frac{1}{N} \sum_{n=1}^{N}
    \mathbbm{1}\left[
        \operatorname{rank}_n(i_n^*) \leq k
    \right]~,
\end{align}
where $i_n^*$ denotes the annotated flare-producing~AR and $\operatorname{rank}_n(i_n^*)$ its rank according to the attention weights.
We additionally analyze the distribution of attention weights assigned to flare-producing and non-producing ARs.
Further details including explicit metric formulae and comparable source identification works are provided in Appendix~\ref{appendix: evaluation}.

\begin{figure}
    \begin{minipage}[t]{.49\textwidth} %0.49
        %\subfloat{
            \stackinset{l}{0cm}{t}{0cm}{\textbf{a}}{
            \includegraphics[width=\textwidth]{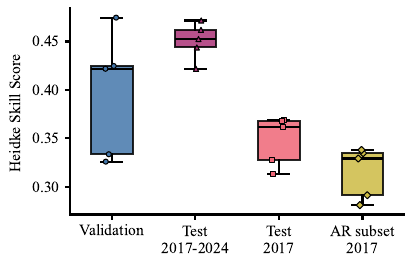}
            }
        %}
    \end{minipage} 
    \hfill
    \begin{minipage}[t]{.49\textwidth}
        
        %\subfloat{
            \stackinset{l}{0cm}{t}{0cm}{\textbf{b}}{
            \includegraphics[width=\textwidth]{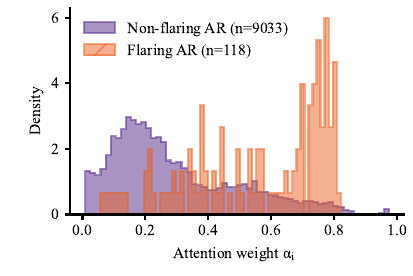}
            %}
        }
        \vspace{6pt}
    \end{minipage}
    \caption{
    \textbf{(a)} Heidke Skill Score~(HSS) across the five temporal cross-validation folds, evaluated on the corresponding validation sets and on three test sets:
    the full 2017--2024 test period, the 2017 subset, and the 2017 subset containing flares with confirmed AR~origins.
    \textbf{(b)} Density-normalized distribution of attention weights assigned to flare-producing and non-producing ARs for the 2017~subset with confirmed AR~origins, using the model with the highest validation~HSS among the five cross-validation models.
    } 
    \label{fig:results}
\end{figure}

\section{Results}
\label{sec: results}
\paragraph{Flare forecasting on future data.}
We first evaluate ARENA on the temporally held-out test set from 2017--2024.
Figure~\ref{fig:results}a shows the HSS obtained by the five models trained on different temporal cross-validation folds, evaluated on both their respective validation sets and the common test sets.
The models exhibit varying performance across the validation folds which could be related to varying class balances across the folds (see Appendix~\ref{appendix: additional results}).
Since the 2017--2024 test set contains significantly more data than the validation folds, we consider it to provide a more robust result.
This test set shows a median~HSS of~$0.45$, slightly higher than the median HSS on the validation data (HSS$=0.42$).
The AR identification for flares can only be evaluated on a subset of the 2017 data due to limited flare-AR labels~\citep{SWAN_SF_flare_AR_dataset}. 
Therefore we show the~HSS for the full 2017 data as well as the specific subset where AR~region labels are available.
ARENA performs comparably to the AR-level supervised baseline (Table~\ref{tab: per-AR baseline}), suggesting that the attention-based aggregation mechanism introduces little performance degradation and that further improvements may instead be limited by the AR~encoder.
\begin{table}[!ht]
    \centering
   \caption{\label{tab: per-AR baseline} Comparison of per-AR instance flare level prediction vs bag level prediction on the same subset of 2017. We report the median metrics across the five validation folds.}
    \begin{tabular}{lcccccc}
    \toprule
        ~ & TSS$\uparrow$ & HSS$\uparrow$ & F1$\uparrow$ & Precision$\uparrow$ & TPR$\uparrow$ & TNR$\uparrow$ \\ \midrule
          Per AR (instance level) & \textbf{0.85} & 0.17 & 0.19 & 0.10 & \textbf{0.96} & \textbf{0.90}  \\ 
        ARENA (bag level) & 0.79	& \textbf{0.33} &	\textbf{0.37}	& \textbf{0.23}	& 0.92 &	0.88 \\
        \bottomrule
    \end{tabular}
\end{table}

\paragraph{AR source identification.}
We next assess whether the learned attention weights identify the AR~associated with a subsequent flare.
Excluding observations containing only one~AR for which source localization is trivial, ARENA achieves an Acc@1 of $87.1\pm0.7\%$ and an Acc@2 of $100.0 \pm 0.0\%$ across the five models.
Figure~\ref{fig:results}b shows the corresponding distribution of attention weights for flare-producing and non-producing~ARs for our best run (validation HSS$=0.47$).
Flare-producing~ARs receive systematically higher attention weights, consistent with the quantitative localization results.

\section{Conclusion}
We introduced \textbf{ARENA}, an attention-based multiple-instance learning architecture for interpretable solar flare forecasting. 
Using only bag-level flare labels and AR cutouts, ARENA achieves a median HSS of~0.45 on the temporally held-out 2017--2024 test set and reaches a performance comparable to a model trained on flare-AR labels, while providing an intrinsic AR-level importance signal.
Despite never observing flare-origin annotations during training, ARENA assigns the highest attention weight to the flare-producing AR in $87.1\pm0.7\%$ of cases. 
These results demonstrate that weak supervision can recover meaningful AR-level associations directly through the model's attention mechanism. 
Future work will extend this framework to other solar eruptive phenomena, different attention mechanisms and automated flare-AR attribution.

\begin{ack}
    The authors thank Vincent Berenz for great support in storing and handling large amounts of solar images and Hsiao-Ru Pan for the thoughtful discussions on the computer vision approaches.
    AK thanks the International Max Planck Research School for Intelligent Systems~(IMPRS-IS) for support.
    The computational work for this paper was performed on the Atlas cluster at the Max Planck Institute for Intelligent Systems. 
    AAR and VMP acknowledge financial support from the Agencia Estatal de
    Investigación (AEI) through the Severo Ochoa Centre of Excellence
    accreditation awarded to the Instituto de Astrofísica de Canarias,
    grant CEX2025-001609-S, funded by MICIU/AEI/10.13039/501100011033.
\end{ack}

\FloatBarrier

%%%%%%%%%%% Bibliography %%%%%%%%%%%
\bibliographystyle{abbrvnat_copy}
\bibliography{mybib}

%%%%%%%%%%%% Appendix %%%%%%%%%%%%%%
\FloatBarrier
\newpage
\appendix

\section{Data set and preprocessing}
\label{appendix:data_set}
\begin{figure}
    \begin{minipage}[a]{\textwidth} %0.49
        \subfloat{
            \stackinset{l}{0cm}{t}{0cm}{\textbf{a}}{
            \includegraphics[width=\textwidth]{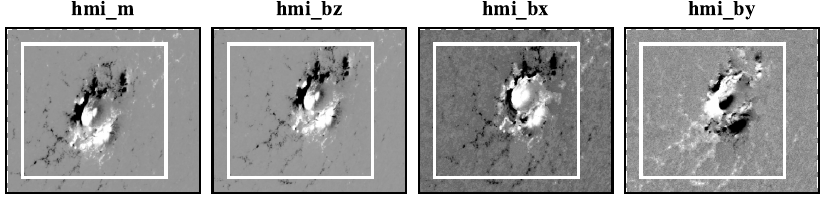}
            }
        }
    \end{minipage} 
    \hfill
    \begin{minipage}[b]{\textwidth}

        \subfloat{
            \stackinset{l}{0cm}{t}{0cm}{\textbf{b}}{
            \includegraphics[width=\textwidth]{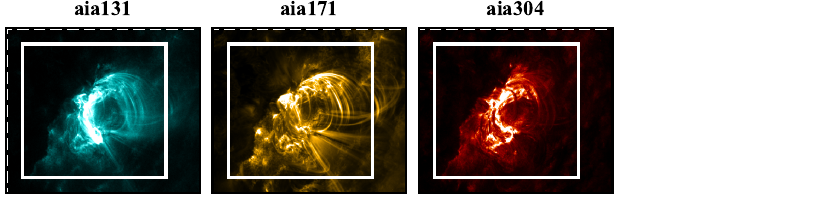}
            }
        }
        \vspace{6pt}
        \caption{SHARP 7115 cutout image HMI and AIA channels at 2017/09/07 01:00    }
        \label{fig: AR channels}
    \end{minipage}
\end{figure}

\begin{figure}
    \begin{minipage}[a]{\textwidth} %0.49
        \subfloat{
            \stackinset{l}{0cm}{t}{0cm}{\textbf{a}}{
            \includegraphics[width=\textwidth]{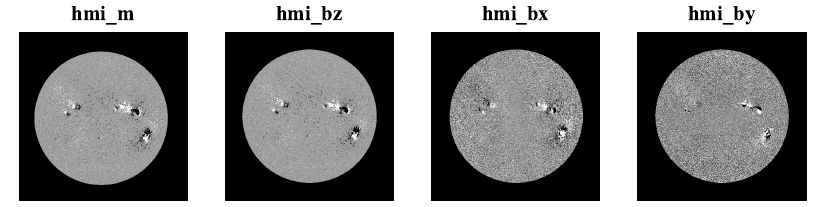}
            }
        }
    \end{minipage} 
    \hfill
    \begin{minipage}[b]{\textwidth}

        \subfloat{
            \stackinset{l}{0cm}{t}{0cm}{\textbf{b}}{
            \includegraphics[width=\textwidth]{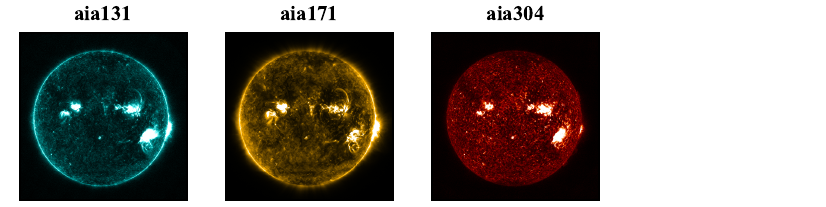}
            }
        }
        \vspace{6pt}
         \caption{Solar full disk images of four HMI and three AIA channels at 2017/09/07 01:00. hmi\_m refers to the line of sight magnetogram.}
         \label{fig: full disk channels}
    \end{minipage}
\end{figure}

\begin{figure}
    \centering
    \includegraphics[width=\textwidth]{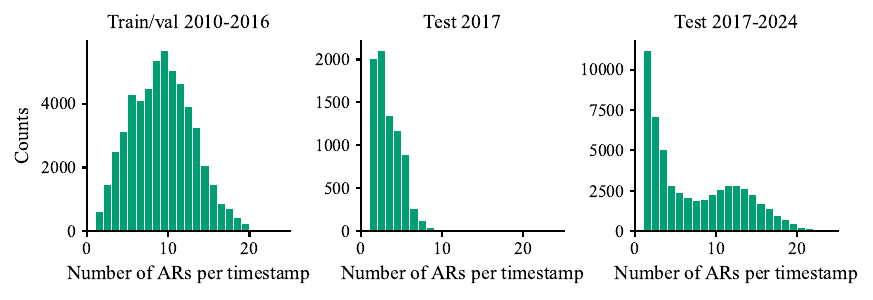}
    \caption{\label{fig: num ARs per timestamp panel} Number of ARs per timestamp across different datasets.}
    \label{fig:placeholder}
\end{figure}

\paragraph{Active region image catalog.}
We represent each solar observation as a variable-sized set of active regions (ARs) rather than using full-disk observations.
ARs are the primary sites of solar flare activity, particularly for large and energetic flares, and therefore provide a physically motivated spatial decomposition of the solar disk.
Restricting the input to AR cutouts also avoids processing the large fraction of the solar disk corresponding to quiet-Sun regions, substantially reducing memory, computational, and storage requirements while retaining the regions most relevant to flare prediction.
\\
We use the Space-weather HMI Active Region Patches (SHARPs) catalog~\citep{hmi_harp_pipeline, Bobra_2014_sharps}, which provides automatically identified and tracked AR patches derived from observations of the Helioseismic and Magnetic Imager~(HMI).
SHARPs are widely used for data-driven solar flare prediction~\citep{Huang_flares_2018, Tang_flares_2021, Sun_flares_2022, Pandey_flares_2024_ARs, Guastavino_2023_flares}.
\\
The SHARPs catalog is derived from observations of the photospheric magnetic field.
A magnetogram measures the magnetic field projected onto a particular direction.
In a line-of-sight~(LOS) magnetogram, the measured quantity is the component of the magnetic field along the line connecting the solar surface to the observer.
In contrast, vector magnetograms provide information about the full magnetic-field vector, including both the LOS and transverse components.
The active-region detection and tracking pipeline used to construct SHARPs operates on line-of-sight and vector magnetograms and identifies regions of enhanced magnetic activity.
In particular, the detection procedure applies Gaussian-kernel filtering to activity bitmaps derived from the magnetograms~\citep{Bobra_2021_SMARPs_and_SHARPs}.
\\
\paragraph{Multi-channel active-region dataset.}
We construct a custom dataset based on the SHARPs catalog by augmenting the magnetic-field information with observations of the solar atmosphere:
Extreme ultra-violet~(EUV) observations provide complementary information about the morphology of the solar corona, including coronal loops and other structures associated with active regions.
In particular, the Atmospheric Imaging Assembly~(AIA) channels \citep{Lemen_2012_AIA} at 131, 171 and 304~\AA\ are each dominated by emission from ions with distinct characteristic formation temperatures, so that the three channels jointly sample plasma across a broad range of coronal and transition-region temperatures. 
This provides morphological information that is not directly captured by photospheric magnetograms alone. Each channel image encodes intensity as a function of position rather than an explicit temperature; we do not perform a differential-emission-measure inversion or otherwise reconstruct a temperature map. 
\\
Instead, by treating the co-aligned channels as separate input channels to the network, we allow the model to exploit channel-specific and cross-channel morphological differences (e.g. a structure that is prominent in one band but faint in another). 
These differences arise from the differing temperature sensitivity of each channel, without requiring the network to represent temperature explicitly.
\\
Recent work has also shown that flare-forecasting models using EUV observations can outperform models based solely on magnetograms~\citep{Patch_cnn_Francisco_2024}.
We also increase the spatial extent of the extracted patches relative to the standard SHARP cutouts, similar to~\citet{Dissauer_2023_AARPs}.
This provides additional context around the active region and captures larger-scale coronal structures, such as extended magnetic loops, that may extend beyond the photospheric AR~boundary.
This way each AR~instance can be viewed as a multi-modal physical representation of the photospheric magnetic field and coronal morphology, with an example visualized in Figure~\ref{fig: AR channels} and the corresponding full-disk images in Figure~\ref{fig: full disk channels}. 
We refer to the resulting multi-channel active-region cutouts as \emph{AIA Active Region Patches}~(AARPs) throughout the remainder of the paper.
\\ 
For extracting the same solar region from AIA and vector magnetogram channels we use the multi-channel full-disk images of the Surya-dataset~\citep{roy2025surya}, covering 05/2010-12/2024 at a $\SI{12}{\min}$ cadence. 
The Surya dataset is based on images from Solar Dynamics
Observatory~(SDO), specifically from the eight AIA~channels~\citep{Lemen_2012_AIA} and five HMI channels \citep{Schou_2012_HMI}. 
The AIA instrument captures EUV and UV images at a cadence of \qtyrange{12}{24}{\second} while the HMI instrument captures images of the magnetic field of the sun at a cadence of 
\SI{12}{\second}.
We use the timestamps provided by Surya at a reduced cadence of two~hours.
This temporal sampling is chosen because the energy build up and large-scale evolution of the magnetic and coronal structures preceding flare events occurs predominantly on timescales of hours to days, while the native instrument cadence would result in substantial redundancy for our forecasting time horizon.
For each selected timestamp, we identify the corresponding ARs and extract spatially aligned multi-channel cutouts from the HMI and AIA observations.
Thus, each timestamp is represented as a variable-sized set of 1-24 AARPs, with an average of approximately 7~AR cutouts per timestamp. The distribution across our datasets can be seen in Figure~
\ref{fig: num ARs per timestamp panel}.
\\

\paragraph{Data normalization and spatial preprocessing.}
The different observational channels exhibit substantially different pixel-value distributions, and we therefore normalize each channel separately.
The EUV intensities are strongly skewed towards higher values, with a distribution that is approximately exponential.
We first apply a logarithmic transformation, $\log(1+x)$, and subsequently standardize each channel to zero mean and unit variance.
In contrast, HMI magnetograms contain both positive and negative values and exhibit a substantial noise-dominated regime around zero, approximately within $\pm7$~Gauss for the line-of-sight channel and $\pm70$~Gauss for the vector magnetograms.
Applying a logarithmic transformation is therefore not appropriate, as it would disproportionately amplify the relative contribution of this noise-dominated regime.
Instead, we clip the magnetogram values to the range $[-300,300]$~Gauss and linearly rescale the resulting values to $[-1,1]$.
The clipping threshold is chosen such that the magnetic structures relevant for flare prediction remain well represented while limiting the influence of extreme values.
The extracted AR patches have variable spatial dimensions.
To obtain a fixed-size input compatible with the ResNet-18 architecture described in Appendix~\ref{appendix: architecture_details}, we resize all patches to $224\times224$ pixels.
This resizing does not preserve the original spatial dimensions of the ARs, so we additionally investigated whether explicit information about the original patch position and size could improve the learned representation.
Specifically, we encoded the original position and dimensions using a separate multilayer perceptron(MLP) and concatenated its output with the ResNet-18 embedding.
This additional positional and scale information did not yield a significant improvement in forecasting performance and was therefore omitted from the final model.

\begin{figure*}[t]
    \centering
    \includegraphics[width=\textwidth]{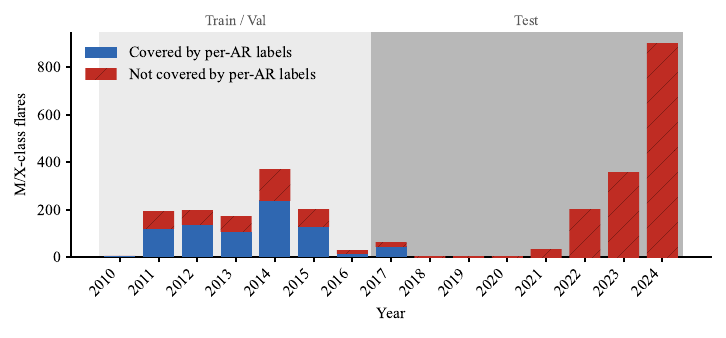}
     \caption{ \label{fig: flare occurences per year}
    Number of flares per year and dataset split. We use the data from 2010-2016 for training and validation. The data from 2017-2024 is used as the holdout test-set. On a subset of the 2017 test data (blue) we evaluate the attention weights for each region as this overlaps with our flare catalog with AR labels~\citep{SWAN_SF_flare_AR_dataset}.
    } 
\end{figure*}

\paragraph{Flare labels.}
Flares are categorized into A, B, C, M, or X-categories according to the peak magnitude of the soft X-ray flux.
%(wavelengths of 0.1--0.8 nm)
We restrict our analysis to M- and X-class flares (M+), corresponding to a peak flux of at least $10^{-5}$~W/m$^2$. 
Energetic events are particularly relevant for space-weather applications because their electromagnetic radiation can affect the Earth's ionosphere and disrupt radio communication and space-based technologies. 
The temporal distribution of flares over the period of our dataset can be seen in Figure~\ref{fig: flare occurences per year}. 
\\
We obtain flare start times and classifications from the NOAA GOES XRS flare catalog~\citep{GOES_XRS_flare_catalog}.
It is based on Observations of the Geostationary Operational Environmental Satellite system~(GOES), specifically the X-Ray Sensor~(XRS). 
\\
We label a selected observation timestamp as positive if at least one M/X-class flare occurs in subsequent \SI{24}{\hour}, otherwise the bag is assigned a negative label. 
Therefore, the forecasting task is to predict wether a M+ flare will occur within the next \SI{24}{\hour} solely based on the ARs present at the observation time.
For training and validation we use the data from 05/2010--11/2016 and reserve the period from 2017--2024 as a  held-out test set. 
Within the training and validation set we perform a five fold temporal cross-validation: 
For each fold, two calendar months from each year are assigned to the validation set (e.g. January and February). 
The two months immediately adjacent serve as temporal buffer, while all the remaining months are used in the training set. 
For example, when January and February are used for validation, December of the preceding year and March are excluded and April to November are used for training.
The temporal buffer reduces the risk of information leakage caused by the strong temporal correlation between neighboring observations and by the persistence of active regions across multiple days and weeks.
The 2017--2024 period is not used for model selection and serves as a future test set, allowing us to evaluate wether the learned representation generalizes to observations from a later time period. 
\\
Because reliable associations between individual flares and their source ARs are available only for a limited subset of the data, we do not use AR-level flare-origin annotations to construct the training, validation, or test labels.
Instead, all labels are derived exclusively from the global GOES flare catalog, consistent with the MIL formulation described in Section~\ref{sec: ARENA}.
The resulting class distributions differ substantially between the temporal splits; detailed class-balance statistics and additional evaluation results are provided in Appendix~\ref{appendix: additional results}.

\paragraph{Flare origin.}
To evaluate whether the learned instance weights identify the AR responsible for a predicted flare, we additionally use the SWAN space-weather dataset~\citep{SWAN_SF_flare_AR_dataset}, which provides associations between individual flare events and their source active regions. 
The dataset is based on GOES XRS flare observations and combines information from multiple flare and AR catalogs to assign an AR origin to individual events. 
The resulting flare-AR associations are available from 2010 through 2017.
We match the flare events in the SWAN catalog to the events in the NOAA GOES XRS catalog using a temporal tolerance of $\pm1$~hour. 
With this tolerance we cover approximately 65\% of flares observed during this period from the GOES XRS catalog. 
For source region identification evaluation and the encoder comparison, we discard the timestamps that have a subsequent flare eruption according to~\citep{GOES_XRS_flare_catalog} but no AR source annotation according to~\citep{SWAN_SF_flare_AR_dataset}.  
There are no flares originating from different ARs within~24h in the relevant time-frame according to the SWAN catalog. 
This means that each of our positively labeled timestamps for this evaluation set can clearly be attributed to one AR.
\\
Importantly, these source-region annotations are used only for evaluation and are not used during training. Because the availability of reliable flare--AR associations is limited to this subset, we restrict the evaluation of the learned AR-level weights to the subset of M+~flares in 2017 for which a source-region annotation is available. 
This means the period of 2017 can be used for bag level flare prediction evaluation and also provides an independent set of flare-AR associations against which the attention weights can be evaluated.
That way, we can assess whether the AR receiving the highest model weight corresponds to the independently annotated flare-producing AR, despite the model having been trained exclusively with bag-level flare labels.
\\
The number of flares per year, in addition to and the availability of corresponding AR origin annotations, can be seen in Figure~\ref{fig: flare occurences per year}. 
Usually humans experts can visually inspect a series of images to decide which AR a flare is originating from. 
Such manual associations are labor-intensive and become increasingly challenging during periods of high solar activity when multiple ARs are simultaneously present.
The SWAN catalog provides an automated alternative based on the cross-matching of multiple observational catalogs, but its incomplete coverage motivates restricting source-region evaluation to the annotated subset.

\section{Architecture details}
\label{appendix: architecture_details}

\paragraph{Encoder network.}
We choose a standard ResNet-18~\citep{ResNet} as an encoder to for the AR image due to its versatility and low parameter count, as we only have access to limited data. 
We change the number of input channels in the first convolutional layer to accommodate all the seven channels from \ref{appendix:data_set}.
This ResNet-18 has 11.2 million parameters in our seven channel configuration. 

\paragraph{Attention-based aggregation.}
As described in Section~\ref{sec: ARENA}, ARENA represents each observation at time~$t$ as a bag of $N_t$ active-region instances with corresponding embeddings $\mathbf{h}_i \in \mathbb{R}^{M}$.
The attention-based multiple-instance learning (ABMIL) aggregator computes a scalar attention score for each instance and uses the normalized scores to form a weighted sum of the instance embeddings:
\begin{equation}
    \mathbf{z}
    =
    \sum_{i=1}^{N_t} \alpha_i \mathbf{h}_i,
    \label{eq: weighted_sum}
\end{equation}
where the attention weights are already normalized via the softmax function in order to make the magnitude of the aggregation vector~$\mathbf{z}$ independent of the number of instances inside a given bag.
\\
For our main ARENA model, we use the gated attention mechanism proposed by~\citet{ilse2018_abmil}.
For each instance embedding $\mathbf{h}_i \in \mathbb{R}^M$, a non-normalized attention score for a give embedding $\mathbf{h_i}$ is computed as
\begin{equation}
    a_{\phi}(\mathbf{h_i})
    =
    \mathbf{w}^{\top}
    \left[
    \tanh\left(\mathbf{V}\mathbf{h}_i^{\top}\right)
    \odot
    \operatorname{sigm}\left(\mathbf{U}\mathbf{h}_i^{\top}\right)
    \right],
    \label{eq:gated_attention_score}
\end{equation}
Here, $\mathbf{V},\mathbf{U}\in\mathbb{R}^{L\times M}$ and $\mathbf{w}\in\mathbb{R}^{L}$ are learnable parameters, $M$ is the dimension of the instance embedding, $L$~is the dimension of the attention space, and $\odot$ denotes element-wise multiplication.
The functions $\tanh(\cdot)$ and $\operatorname{sigm}(\cdot)$ denote the hyperbolic tangent and sigmoid nonlinearities, respectively.
The gating mechanism combines two independently transformed versions of the instance embedding through element-wise multiplication before projecting the result to a scalar attention score.
This allows the attention function to model more complex relations rather than relying on a single nonlinear transformation of the embedding.
The resulting aggregation is invariant to the ordering of the active regions and naturally accommodates observations containing different numbers of ARs. 
The magnitude of the aggregated vector $\mathbf{z}$ is also independent of the sequence length due to the softmax normalization of the weights. 
\\
For our experiments, the ResNet-18 encoder produces embeddings of dimension $M=512$ and we use an attention dimension of $L=128$.
The gated attention module therefore contains
\begin{equation}
    2LM + L = 2 \cdot 128 \cdot512 +128 = 131{,}200,
\end{equation}
trainable parameters, excluding any parameters associated with the ResNet-18 encoder and downstream classification head.
\\ 
The aggregation vector~$\mathbf{z}$ is then passed into a small classifier network which consists of two linear layers, projecting the 512-dimensional down to 256 dimension, applying the ReLU activation function and projecting into a single logit for classification. This small classifier network has $512 \cdot 256 +256 \cdot 1 =  131{,}328$ parameters.

\section{Training details}
\label{appendix:training_details}
Since M+ events constitute only a minor fraction of Sun observations, the forecasting dataset is strongly imbalanced. 
We therefore investigated several strategies for addressing class imbalance during training, including negative-instance undersampling, weighted random sampling, and class-weighted loss functions.
Among these approaches, undersampling the negative training observations to obtain a balanced training set yields the best validation performance.
Undersampling is applied only to the training data; the validation and test sets retain their original class distributions.
Without undersampling, we observe a strong tendency toward the majority-class solution, with models assigning negative predictions to nearly all observations.
Weighted random sampling and using a class-weighted loss reduce this effect but do not provide a consistent improvement over negative undersampling.
\\
Therefore, we use negative undersampling for all experiments reported in this paper.
We selected the optimization hyperparameters based on the validation performance during the development period.
The final configuration uses a learning rate of $\eta=10^{-7}$, a weight decay of $0.001$, and a batch size of~32.
Higher learning rates resulted in rapid degradation of validation performance, consistent with overfitting.

\section{Evaluation}
\label{appendix: evaluation}
\paragraph{Binary prediction metrics.} We evaluate binary M+ flare forecasting using the True Skill Statistic~(TSS), Heidke Skill Score~(HSS), precision and F1-score.
All three metrics are computed from the confusion matrix consisting of true positives~(TP), false positives~(FP), true negatives~(TN), and false negatives~(FN).
Unless stated otherwise, the predicted class is obtained by applying the same decision threshold of $0.5$ to the model output probability on the evaluation sets.
\\
The TSS is defined as
\begin{equation}
    \label{eq: tss}
    TSS = TPR + TNR -1 =\frac{TP}{TP + FN} + \frac{TN}{TN+FP} -1 \in [-1, 1],
\end{equation}
where $TPR$ and $TNR$ denote the true-positive rate (recall) and true-negative rate (specificity), respectively. 
TSS ranges from $-1$ to $1$, with $1$ corresponding to perfect classification and $0$ corresponding to no discriminatory skill under the standard interpretation.
Unlike accuracy, TSS is insensitive to changes in the class prevalence because it gives equal weight to the true-positive and true-negative rates.
In theory, this makes it particularly suitable for evaluating flare forecasting under strong class imbalance~\citep{Flare_prediction_metrics_bloomfield}.
\\
The HSS measures the improvement in classification performance relative to the expected performance of random classification given the observed marginal class frequencies.
We compute it as
\begin{equation}
    \label{eq: HSS}
    HSS = \frac{2 (TP \cdot TN - FP \cdot FN)}{(TP + FN)  (FN + TN) + (TP + FP)  (FP + TN)} \in [-1, 1].
\end{equation}
An HSS of~$1$ indicates perfect classification and $0$ indicates no improvement over the corresponding chance agreement.
Unlike TSS, HSS depends on the class distribution and can therefore be affected by changes in event prevalence between datasets.
\\
Additionally,  we report the precision, defined as
\begin{equation}
    \label{eq: precision}
    Precision = \frac{TP}{TP+FP} \in [0, 1].
\end{equation}
Precision measures the fraction of predicted flares that correspond to actual flares. It therefore directly quantifies the reliability of positive predictions: low precision indicates a high rate of false alarms among predicted flare events. 
\\
The F1 score is the harmonic mean of the TPR and the precision:
\begin{equation}
    F1 =  2 \cdot \frac{Precision \cdot TPR}{Precision + TPR} = \frac{2 TP}{2TP +FP +FN} \in [0, 1].
\end{equation}
The F1 score also ranges from one, indicating perfect precision and recall, to zero if either precision or TPR is zero.

\paragraph{Instance-level encoder baseline.}
The instance-level classifier, whose performance is reported in Table~\ref{tab: per-AR baseline} in Section~\ref{sec: results}, uses the same ResNet-18 encoder and classification MLP as ARENA to predict M-class flares from individual AR cutout images. 
Unlike ARENA, which learns from bag-level event labels through attention-based aggregation, the encoder baseline is trained using explicit AR–flare associations from the SWAN catalog~\citep{SWAN_SF_flare_AR_dataset}. Both models are trained and evaluated using the same temporal splits.
\\
A direct comparison of their performance metrics must account for the different class prevalences in the two datasets:
Although both datasets contain the same number of positive examples, these examples have different levels of granularity: 
positive AR instances in the instance-level dataset correspond to positive bags in the ARENA dataset. 
In the instance-level dataset, each AR cutout is labeled individually, whereas ARENA assigns a single label to the entire bag of AR cutouts. 
Consequently, AR instances in a positive bag that are not identified as flare-associated on an instance-level contribute as negative examples to the instance-level dataset, increasing the number of negative examples for the instance-level classifier. 
As a result, positive instance examples account for only 1.2~\% of the instance-level dataset, compared with 3.86~\% positive bags of the ARENA dataset.
Consequently, the comparison should not be interpreted as a fully controlled benchmark as HSS, F1, and precision depend heavily on the class composition of the dataset.
Nevertheless, it provides an indication of whether the weakly supervised aggregation introduces a substantial performance limitation with regard to the TSS, TPR, and TNR metrics.
\\

\paragraph{Source region identification.}
As described in Section~\ref{sec: experimental setup}, we evaluate whether the attention-based ranking of active regions is consistent with the independently annotated flare source. 
For each positive observation, the ARs are ranked according to their normalized attention weights $\{\alpha_i\}$, and the resulting ranking is compared with the annotated source ARs recorded in the SWAN flare-AR catalog \citep{SWAN_SF_flare_AR_dataset}.
\\
This evaluation is restricted to the temporally held-out 2017 test subset for which flare-AR associations are available, see Appendix~\ref{appendix:data_set}. 
For this specific subset of 2017, we discard all positive timestamps where the source region is unknown according to the SWAN catalog but keep all the negatives. 
The flare-AR annotations are not used during training, model selection, or attention weight computation. 
They are used exclusively for this post-hoc evaluation. 
Thus, source-region identification provides an independent assessment of whether the weakly supervised attention mechanism assigns high importance to the AR associated with the subsequent flare.
\\
Identifying the flare origin is also a goal of the work by~\citet{Patch_cnn_Francisco_2024} which introduced Patch-CNN~(P-CNN). 
There, the authors cut the full disk image into fixed rectangular tiles, apply a flare prediction network on the individual tiles and then use the maximum output probability over the different tiles to train their network on global flare prediction.
The authors use EUV brightness center-of-mass calculations during a flare to identify its exact source region and compare it to post-hoc GradCam visualizations to evaluate their source region identification.
However, the specific dataset and chosen evaluation period do not allow us to make a detailed comparison which will be explained in detail when the baseline approaches are introduced.
\citet{pandey_2023_towards_interpr} used attention modules on the feature maps of different convolutional blocks. 
However, they only evaluate individual cases qualitatively without a global quantitative evaluation. 
\citet{Pandey_flares_2024_ARs} and \citet{Kangwo_2021_first_grad_cam} used GradCam based methods to locate relevant features post-hoc after training the network for flare prediction without including a quantitative evaluation.

\paragraph{Flare prediction baselines.}
We compare ARENA results with three baselines: We (1)~train a full-disk model from scratch, (2)~fine-tune the Surya foundation model, and (3)~report the results of P-CNN~\citep{Patch_cnn_Francisco_2024}.
As a representative and directly comparable baseline, we train a full-disk model using the same seven observational channels as ARENA. 
We spatially downsample the $4096\times4096$ full-disk observations from the Surya dataset to $224\times224$ pixels and process them with the same ResNet-18 encoder~\citep{ResNet} and classification MLP used by ARENA. 
This provides a direct comparison between the proposed representation as a variable-sized set of high-resolution AR cutouts and a single low-resolution full-disk representation while keeping the image encoder and classification head fixed.
\\
Additionally, we fine-tune the Surya foundation model~\citep{roy2025surya} for M+ flare forecasting.
Surya is pretrained using a pixel-level next-image prediction objective on multiple solar-observation channels, including the EUV channels used in our input representation. 
Because EUV observations can contain signatures of solar flares, evaluating on observations that were included in the Surya pretraining data would result in pretraining exposure to the evaluation period, resulting in data leakage. 
We therefore restrict evaluation to observations that were not used during Surya pretraining.
\\
In the original pretraining procedure, days 16–30 of each year from 2010–2019 are used for validation, days 1–15 and 31–45 serve as temporal buffers, and the remaining observations before 2020 are used for pretraining. 
Observations from 2020–2024 constitute the held-out period. 
For our downstream flare-prediction experiment, we retain this temporal structure but restrict supervised training to observations from 2010–2016 and also undersample the negatives during training to match the training configuration for ARENA as closely as possible.
We subsequently evaluate the fine-tuned model on observations from 2020–2024.
\\
Finally, we compare against P-CNN~\citep{Patch_cnn_Francisco_2024} on flare prediction.
P-CNN divides each full-disk solar observation into rectangular patches, processes each patch independently with a classifier, and uses the maximum patch-level probability as the global flare prediction. 
Since the authors do not provide an implementation, we use the performance values reported in their publication rather than re-training a model ourselves. 
They train separate models on three EUV channels and one the line-of-sight HMI magnetograms on data from 2010-2019 and report results on the test set from January 2020 through April 2023.
We use the results of their best model~(EUV), and compare our models on the corresponding period for comparison.
They report their cross-validation results as the results of one ensemble model by averaging the probability outputs of their five models trained on different cross-validation folds. 
Therefore we do the same for our ARENA and full-disk model. 
For Surya this is not feasible due to aforementioned data leakage and limited computational resources.
\\
The low-resolution full disk baseline is the most relevant to ARENA, as the results of Surya and Patch-CNN are not directly comparable to ARENA due to the different data splits and pre-processing procedures. 
For our Surya and low-resolution full-disk implementation, we only used the timestamps for which there are ARs according to \citep{Bobra_2014_sharps} in order to replicate the dataset statistics. 
We compare the models on the same subset of 01/2020--04/2023 as the metrics change significantly across different datasets with different class balances which we will discuss later in Appendix~\ref{appendix: additional results}.

\paragraph{Additional evaluation analyses.} In addition to the primary metrics reported in the main text we perform several analyses to characterize model performance under different evaluation conditions. 
We report confusion matrices and additional classification metrics for the validation folds and the different temporally held-out test subsets. 
We further investigate the relationship between class prevalence and the resulting performance metrics. 
Finally we look into our best validation fold (September/October, HSS=0.47) on the test set and analyze the most prevalent failure mode, false positives, in more detail.

\section{Additional results }
\label{appendix: additional results}
\begin{figure*}[!ht]
    \centering
    \includegraphics[width=\textwidth]{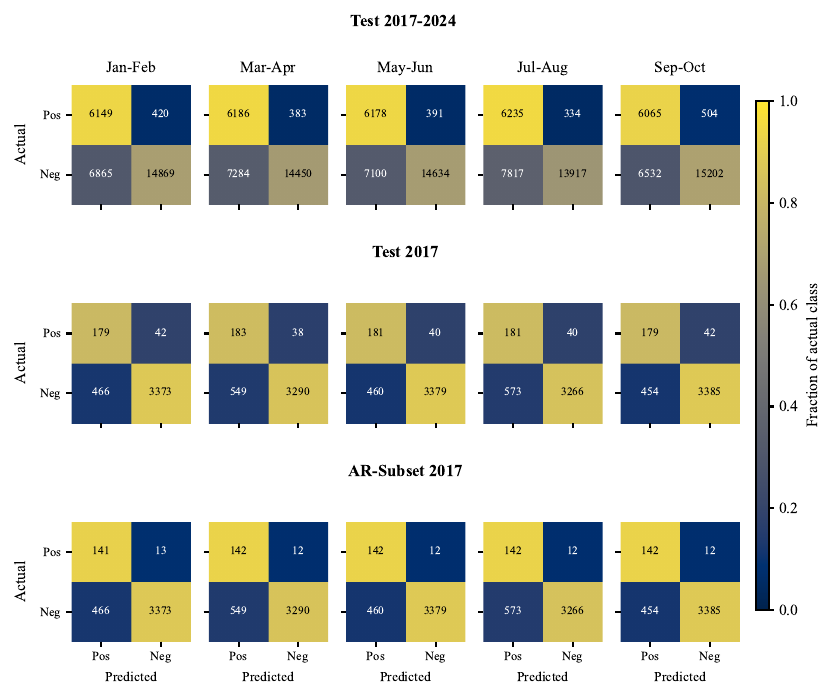}
    \caption{ \label{fig: confusion matrices test set}
    Confusion matrices for the different temporally held-out test subsets. 
    The model correctly identifies a majority of both positive and negative observations, with consistent performances across each evaluation set.
    This can be observed when comparing TP~and~FN as well as TN~and~FP values, the model exhibits good predictive performance.
    This is also the case for the ratio of FN~and~TN values.
    However, comparing TPs~and~FPs, we observe that the model predicts a large number of FP values, especially in the 2017 test sets.
    The background colors are normalized per row of each matrix.
    } 
\end{figure*}

\begin{figure}
    \begin{minipage}[b]{.49\textwidth}
        \subfloat{
            \stackinset{l}{0cm}{t}{0cm}{\textbf{a}}{
            \includegraphics[width=\textwidth]{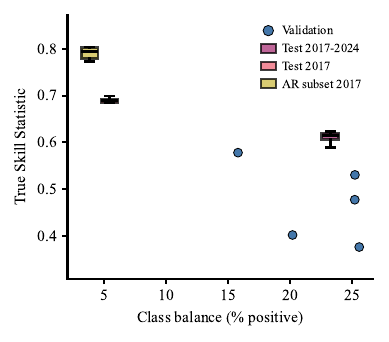}
            }
        }
    \end{minipage} 
    \hfill
    \begin{minipage}[b]{.49\textwidth}
        
        \subfloat{
            \stackinset{l}{0cm}{t}{0cm}{\textbf{b}}{
            \includegraphics[width=\textwidth]{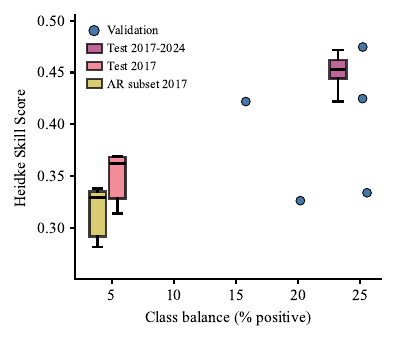}
            }
        }
        \vspace{6pt}
    \end{minipage}
    \caption{ \label{fig: class balance TSS and HSS}
        (a) TSS vs.\ positive-class prevalence:
        The evaluation subsets with lower positive-class prevalence tend to exhibit higher TSS, with the 2017 and 2017 AR-annotated subsets reaching values of approximately 0.69--0.80, compared with 0.40--0.62 for the other subsets.
        (b) HSS vs.\ positive-class prevalence:
        HSS shows a weaker tendency in the opposite direction, with values of approximately 0.29--0.42 for the highly imbalanced 2017 subsets and approximately 0.45 for the full test set. The validation subsets show no clear monotonic relationship.
    }
\end{figure}

To contextualize the results shown in Figure~\ref{fig:results}a, we first discuss the confusion matrices across the different validation folds on the test set and its subsets in Figure \ref{fig: confusion matrices test set} and provide additional metrics in Table \ref{tab: Additional performance metrics}. 
Secondly, we look into the relationship between class balance and the resulting performance metrics in Figure \ref{fig: class balance TSS and HSS}. 
\paragraph{Flare prediction and cross-validation.}
The results across the five cross-validation folds and the temporally held-out test subsets are summarized in Table~\ref{tab: Additional performance metrics}. 
The corresponding confusion matrices for the test set and its subsets are shown in Figure~\ref{fig: confusion matrices test set}. 
Across the evaluation subsets, ARENA maintains relatively high true-positive and true-negative rates, indicating that the model correctly identifies a substantial fraction of both flaring and non-flaring observations at the selected operating point.
The ratio of TP to FP is reasonable for the 2017--2024 test set, but decreases for the 2017 tests sets.
This behavior is expected for rare-event forecasting: 
even with good TPR and TNR, the precision~(Equation \ref{eq: precision}) can be low due to a large class imbalance.
Consequently, the model can achieve a high TSS~(Equation \ref{eq: tss}) while simultaneously exhibiting relatively low precision on the naturally imbalanced test distribution. 
\\
Interestingly, when excluding the positive timestamps without AR source labels as identified by~\citep{SWAN_SF_flare_AR_dataset} from the 2017 test set (see Figure~\ref{fig: confusion matrices test set}, going from the middle row to the bottom row), we observe a reduction of approximately 20\% in true positives (top left matrix entries) and 70\% in false negatives (top right matrix entries).
This indicates  that the model has problems in the classifying flares where there is AR label according to our catalog~\citep{SWAN_SF_flare_AR_dataset}.
Potentially, these could be quiet Sun flares which do not originate from an AR, but further investigations with a bigger sample size are needed to draw firm conclusions.

\paragraph{Extra scalar parameters.}
We additionally evaluate whether incorporating the SHARP magnetic summary parameters~\citep{Bobra_2014_sharps} or the spatial position and size of each AR into the embedding helps in flare prediction.
Neither modification results in a consistent improvement over the image-only ARENA model and all results are reported for the image-only ARENA model.

\begin{table}[!ht]
    \centering
    \caption{ \label{tab: Additional performance metrics}
    Additional performance metrics on the different evaluation sets in the format $median^{Q75}_{Q25}$.
        } 
    \begin{tabular}{lcccccc}
    \toprule
        Dataset  & TSS$\uparrow$ & HSS$\uparrow$ & F1$\uparrow$ & Precision$\uparrow$ & TPR$\uparrow$ & TNR$\uparrow$ \\ \midrule
         Validation & $0.48^{+0.05}_{-0.08}$ & $0.42^{+0.00}_{-0.09}$ & $0.54^{+0.05}_{-0.01}$ & $0.46^{+0.05}_{-0.04}$ & $0.70^{+0.04}_{-0.05}$ & $0.77^{+0.02}_{-0.02}$  \\ 
         Test 2017--2024 & $0.61^{+0.01}_{-0.01}$ & $0.45^{+0.01}_{-0.01}$ & $0.62^{+0.01}_{-0.01}$ & $0.47^{+0.01}_{-0.01}$ & $0.94^{+0.00}_{-0.00}$ & $0.67^{+0.01}_{-0.01}$  \\ 
         Test 2017 & $0.69^{+0.00}_{-0.00}$ & $0.36^{+0.01}_{-0.03}$ & $0.41^{+0.01}_{-0.03}$ & $0.28^{+0.00}_{-0.03}$ & $0.82^{+0.00}_{-0.01}$ & $0.88^{+0.00}_{-0.02}$  \\
        AR subset 2017 & $0.79^{+0.01}_{-0.02}$	& $0.33^{+0.01}_{-0.04}$ &	$0.37^{+0.01}_{-0.03}$	&$0.23^{+0.00}_{-0.03}$	& $0.92^{+0.00}_{-0.00}$ &	$0.88^{+0.00}_{-0.02}$ \\
        \bottomrule
    \end{tabular}
\end{table}

\paragraph{Baseline comparison.}
The baseline performance on the relevant subset of our test set can be seen in Table~\ref{tab: P-CNN and Surya baseline}.
Our ARENA model performs similar to our full disk baseline across most metrics, but it does not reach the performance of P-CNN and Surya. 
Interestingly enough, ARENA has the highest TPR. 
ARENA closely matching the performance of the full disk network suggests that no information is lost when only focusing on the AR cutouts instead of the full disk images. Due to the different training splits P-CNN and Surya are not ideally controlled comparisons. 
Surya also has roughly 20x more parameters and therefore even fine-tuning with undersampling requires significantly more resources.
\begin{table}[!ht]
    \centering
    \caption{\label{tab: P-CNN and Surya baseline}
        Baseline comparison of ARENA on the January 2020--April 2023 subset of the test period, corresponding to the evaluation period reported for P-CNN~\citep{Patch_cnn_Francisco_2024}. 
        ARENA and the full-disk ResNet, 
        were implemented and trained by the authors; P-CNN results are taken from the original paper~\citep{Patch_cnn_Francisco_2024}.
        Pre-processing and training protocols differ between methods and therefore the comparison should not be interpreted as a fully controlled benchmark. 
        Detailed experimental configurations are provided in Appendix~\ref{appendix: evaluation}.
    }
    \begin{tabular}{llcccccc}
    \toprule
        ~ &  Training set  & TSS$\uparrow$ & HSS$\uparrow$ & F1$\uparrow$ & Precision$\uparrow$ & TPR$\uparrow$ & TNR$\uparrow$ \\ \midrule
        P-CNN (ensemble)~\citep{Patch_cnn_Francisco_2024} & 2010--2019  & \textbf{0.58} & 0.42 & 0.53 & 0.39 & 0.82 & 0.76  \\ 
         Full disk (ensemble) & 2010--2017 & 0.52 & 0.34 & 0.49 & 0.34 & 0.81 & 0.71 \\
        Fine-tuned Surya~\citep{roy2025surya} & 2010--2017 & 0.48 & \textbf{0.46} & 
        \textbf{0.55} & \textbf{0.54} & 0.57 & \textbf{0.90} \\ 
         ARENA (ensemble) & 2010--2017 & 0.51 & 0.30 & 0.46 & 0.31 & \textbf{0.87} & 0.64 \\\bottomrule
    \end{tabular}
\end{table}

\paragraph{Metric dependence on class imbalance.}
The evaluation subsets differ substantially in their proportion of positive observations, which provides an important confounding factor when comparing performance across datasets. 
We therefore analyze the relationship between the positive-class prevalence and the reported classification metrics for the five validation folds, the full test set, and its subsets. The results for TSS and HSS are shown in Figure~\ref{fig: class balance TSS and HSS}.
\\
We observe a pronounced increase in TSS for the more strongly imbalanced evaluation subsets. In particular, the 2017 and 2017 AR-annotated subsets achieve high TSS values over 0.69, compared to significantly lower values for the more balanced validation and full test sets. This behavior is not expected from class prevalence alone, since TSS is defined in terms of TPR and TNR and is theoretically prevalence-independent. 
This could mean that the operating characteristics of our model, particularly the true-positive and false-positive rates, vary across datasets with different class prevalences or that our model simply performs better on the 2017 evaluation sets. 
The high TSS observed for the 2017 AR subset should still be interpreted with caution, as it is estimated from a comparatively small number of positive observations.
\\
For HSS, we observe a weaker tendency in the opposite direction: the highly imbalanced 2017 subsets achieve HSS values of approximately 0.29--0.37, whereas the full test set reaches approximately~0.45. 
The validation folds do not exhibit a clear monotonic relationship. These observations demonstrate that the choice of evaluation metric can substantially affect the apparent performance when the event prevalence differs between evaluation sets.

\begin{figure}
    \subfloat{
            \stackinset{l}{0cm}{t}{0cm}{\textbf{a}}{
            \includegraphics[width=\textwidth]{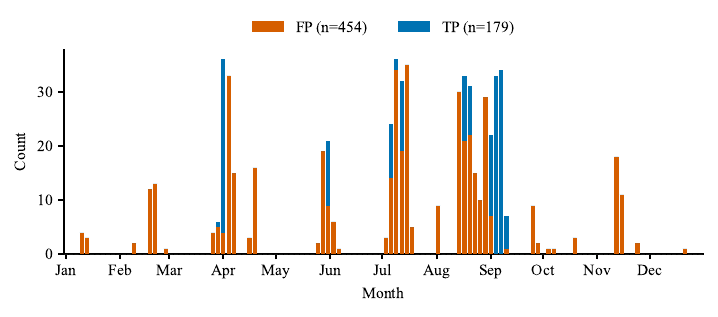}
            }
        }
    \\
    \subfloat{
            \stackinset{l}{0cm}{t}{0cm}{\textbf{b}}{
            \includegraphics[width=\textwidth]{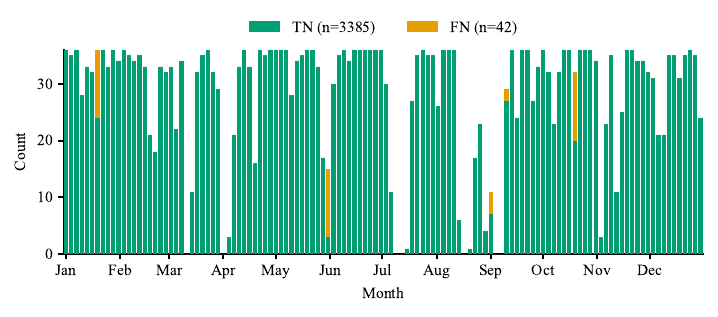}
            }
        }
    \caption{\label{fig: confusion matrix 2017 time histogram} 
        Temporal distribution of the four binary classification outcomes for the 2017 test subset.
        True positives and false positives are shown among the predicted-positive observations, while true negatives and false negatives constitute the predicted-negative observations. 
        False positives frequently occur in temporal proximity to true positives.
    }
\end{figure}

\begin{figure}
    \centering
    \includegraphics[width=\textwidth]{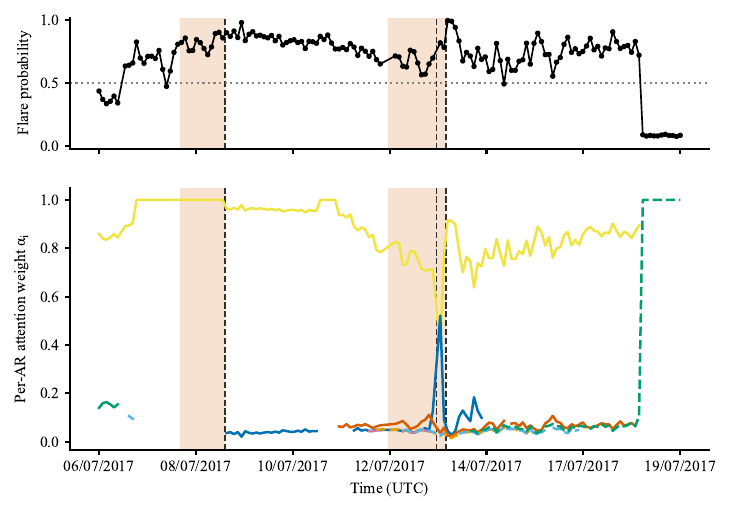}
    \caption{\label{fig: prob and weights time series}Temporal evolution of ARENA's predicted M+ flare probability (top) and attention weights assigned to individual active regions (bottom) for July 6th--19th, 2017. 
    The orange shading indicates observations labeled positive according to our flare catalog because an M+ flare occurs within the subsequent 24-hour prediction window. Flares are represented by the black vertical dashed lines.  
    The same active region receives high attention weights $\alpha_i$ across multiple consecutive observations surrounding the two flare events, while the predicted flare probability remains elevated over an extended period. As soon as this region disappears on July 18th, the output probability of the model drops. The yellow line corresponds to SHARP 7075 which is actually causing the two flares according to \citep{SWAN_SF_flare_AR_dataset}.
    }
\end{figure}

\paragraph{False-positive analysis.}
We further investigate the relatively large number of false positives observed for the 2017 test subset. 
As shown in the third row of Figure~\ref{fig: confusion matrices test set}, this subset contains approximately three times as many false positives as true positives.
The absolute number of false positives is partly a consequence of the strong class imbalance: even a comparatively small false-positive rate can result in many false-positive predictions when non-flaring observations greatly outnumber flaring observations. 
In addition, ARENA is trained using an approximately balanced sampling distribution, whereas the test set follows the substantially lower natural flare prevalence, resulting in a shift in class priors between training and evaluation.
\\
To investigate whether the false positives exhibit additional temporal structure, we visualize the four classification outcomes over time for the 2017 test subset in Figure~\ref{fig: confusion matrix 2017 time histogram}. 
It can be seen that false positives frequently occur immediately before or after sequences of true-positive observations.
This suggests that a substantial fraction of the false positives are not isolated predictions but occur during periods in which the model assigns consistently high flare probability.
To investigate the failure mode of temporally clustered false positives around the true positives further we visualize the temporal evolution of the output probability and the relevant attention weights of the cluster in early July visible in Figure~\ref{fig: confusion matrix 2017 time histogram}. 
The resulting time-series is show in \ref{fig: prob and weights time series}.
We can clearly see one AR in yellow is dominating the attention. Timestamps leading up to the actual flares overlaid in orange are almost all correctly classified as positives. 
However we have significant false positives before and after the events. 
It appears that ARENA is learning an AR's elevated flare propensity over time, while the binary training target imposes a relatively sharp temporal boundary through the 24-hour prediction window.

\section{Related work}
\label{appendix:related_work}
We already discussed Surya~\citep{roy2025surya} and P-CNN~\citep{Patch_cnn_Francisco_2024} in the introduction of the baselines in section~\ref{appendix: additional results}.
For image-based flare prediction, all approaches either use full disk images~\citep{Pandey_flares_2024_full_disk} or AR~images with flare labels~\citep{Pandey_flares_2024_ARs, Li_flares_2023, Guastavino_2023_flares, Sun_flares_2022, Tang_flares_2021}, as it is well-known that most flares originate from ARs~\citep{Overview_of_solar_flares}.  \citet{Huwyler_2022_flare_MIL} use MIL on different solar spectra to decide which frequency range is potentially informative for flare prediction.

\end{document}